# Thermo-mechanical behaviour of a tungsten first wall in HiPER laser fusion scenarios

D Garoz[1,2], A. R. Páramo[1], A Rivera[1*], J M Perlado[1] and R González-Arrabal[1]

[1]Instituto de Fusión Nuclear, UPM, José Gutiérrez Abascal 2, E28006 Madrid, Spain.
[2]Mechanics of Material and Structures, Ghent University, Tech Lane Ghent Science Park – Campus A, Techonologiepark Zwijnaarde 903, 9052 Zwijnaarde (Gent), Belgium

[*]Corresponding author: antonio.rivera@upm

**Abstract.**

The behaviour of a tungsten first wall is studied under the irradiation conditions predicted for the different operation scenarios of the European Laser fusion project HiPER, which is based on direct drive targets and an evacuated dry wall chamber. The scenarios correspond to different stages in the development of a nuclear fusion reactor, from proof of principle (bunch mode facility) to economic feasibility (pre-commercial power plant). This work constitutes a quantitative study to evaluate the first wall performance under realistic irradiation conditions in the different scenarios. We calculated the radiation fluxes assuming the geometrical configurations reported so far for HiPER. Then, we calculated the irradiation-induced first wall temperature evolution and the thermo-mechanical response of the material. The results indicate that the first wall will plastically deform up to a few microns underneath the surface. Continuous operation in power plant leads to fatigue failure with crack generation and growth. Finally, the crack propagation and the minimum W thickness required to fulfil the first wall protection role is studied. The response of tungsten as first wall material as well as its main limitations will be discussed for the HiPER scenarios.

## 1. Introduction

The energy solution to the ever-growing energy demand is a long standing problem that requires innovative solutions. Nowadays, fusion energy can turn into a real option to fossil fuel, with the advantages of being sustainable and environmentally friendly. Two are the main approaches to fusion energy: magnetic confinement (MC) and inertial confinement (IC) by laser (laser fusion). The most advanced projects to demonstrate the viability of laser fusion energy are LIFE (Laser Inertial Fusion Energy) in U.S.A [1–4] and HiPER (High Power Laser Energy Research) in Europe [5–10].

One of the challenging tasks in IC reactors is the development of First Wall (FW) materials able to withstand the hostile environment (high thermal loads and atomistic damage) during power plant operation. In this context, we study different scenarios, corresponding to different phases in the development of nuclear fusion technologies, experimental facilities (Experimental), reactors operating in a relaxed mode far from full power (Prototype) and full scale reactors (Demo). In the different cases, an assessment for the operation of the tungsten FW is necessary. Different approaches are used to reduce damage in FW materials for evacuated dry wall reaction chambers. Projects like LIFE, designed to operate with indirect drive targets, propose filling the chamber with a residual high-Z gas such as Xe (pressure ~1 mbar at room temperature) in order to mitigate ion irradiation [4]. However, this approach is unfeasible in projects based on direct drive targets, like HiPER, because the residual gas makes impossible to maintain the DT target in solid state during injection [11]. Therefore, HiPER will use a dry wall evacuated chamber, i.e., with no mitigation gas for wall protection.

The main requisite for the First Wall material is to have excellent structural stability since severe cracking or mass loss would hamper their protection role. Additional practical requirements for plasma facing materials are: (i) high thermal shock resistance, (ii) high thermal conductivity, (iii) high melting point, (iv) low physical and chemical sputtering and (v) compatibility with the heat sink structure. Moreover, safety and regulation reasons impose low tritium retention as a must. Due to its properties, tungsten is one of the most attractive materials proposed for FW applications [12]. The thermo-mechanical response of W to intense thermal loads has been extensively studied either experimentally [13–16] or by computer simulations [17–19]. The results indicate that the degradation of W properties under thermal loads is a serious concern for the use of W as FW material. Surface deterioration such as thermal shock-induced crack networks [20, 21], roughening [22], blistering, erosion of the loaded tungsten surface [23, 24], appear well below the melting temperature of W. These effects turn out to be unacceptable for a FW material. In order to account for the effect of pulse duration on the thermomechanical response, the heat flux factor ($F_{HF}$) [15, 24] can be used. The $F_{HF}$ is defined as the product of power surface density ($P_{abs}$) and the square root of the pulse duration ($D_t$), i.e., $F_{HF}=P_{abs} D_t^{0.5}$. For ion irradiation, damage appears at fluences ~1.25 J/cm$^2$ [13] for pulses of ~0.5 μs, which means a threshold $F_{HF}$ ~17.7 MWm$^{-2}$s$^{0.5}$. In the case of X-ray irradiation, it has been demonstrated that fluences up to ~1 J/cm$^2$ [13] do not induce appreciable damage on tungsten.

In this work, we quantitatively evaluate the performance of a tungsten first wall under realistic irradiation conditions in the proposed HiPER scenarios (Experimental, Prototype and Demo). The goal is to determine the operation windows of the FW, needed for the construction phase of HiPER. The paper is structured as follows: (i) after describing the HiPER scenarios, radiation fluxes are characterized in each case; (ii) the radiation-induced temperature enhancement and stress generation are determined for every scenario; (iii) the thermomechanical behaviour, damage and crack propagation and the minimum FW thickness is evaluated.

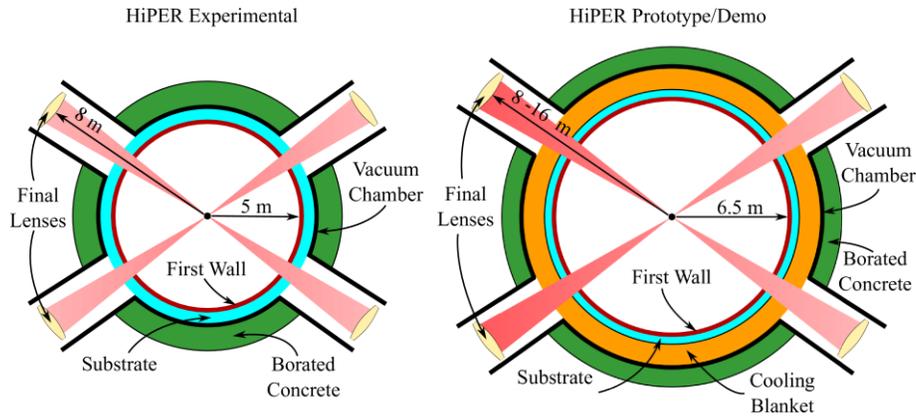

Figure 1: 2D schemes of the laser fusion reactor chambers: (a) HiPER Experimental; (b) HiPER Prototype and Demo.

Table 1: Conditions in the different HiPER scenarios studied in this work: Experimental, Prototype and Demo.

| Parameter | Experimental | Prototype | Demo |
|---|---|---|---|
| Frequency | Few shots per bunch | 1 Hz | 10 Hz |
| Shot Energy (MJ) | 20 | 50 | 154 |
| Chamber radius (m) | 5 | 6.5 | 6.5 |

## 2. Chamber and Radiation Fluxes

Although there is not a definite design for the HiPER chamber, some advanced concepts have been already reported [5, 9]. Different scenarios for HiPER development are foreseen [10]. First, an Experimental facility aimed to demonstrate an advanced ignition scheme and repetitive laser operation; where irradiation is limited to bunches of a few shots with energies of ~20 MJ. Secondly, a Prototype reactor that allows for research on low yield targets, studying target injection, tracking, repetition mode, heat extraction or tritium production while keeping material demands low, operating with relaxed conditions of 1 Hz and 50 MJ. Finally, a Demonstration (Demo) reactor aimed to prove feasibility of technologies under very demanding conditions, operating with high repetition rates of 10 Hz and high energy gains of ~150 MJ. In Table 1 the main conditions of the different HiPER scenarios are shown. In the different scenarios, the chamber is assumed to be spherical, having 48 openings for the laser beam lines, allowing for symmetrical target illumination. Schematic representations of the chambers are shown in Figure 1. The chamber inner radius depends on the facility, being 5 m for the Experimental facility and 6.5 m for the Prototype and Demo reactors. In all cases the chamber has an onion-like shape consisting of different layers. (i) *first wall (FW)*, directly facing the irradiation and made of W [25]; (ii) *substrate,* which gives mechanical support to the FW and it is proposed to be made of ODS-RAFM steel [26]; (iii) *coolant*, for the Prototype and Demo scenarios, the chamber is also surrounded by a thick liquid LiPb *blanket* [9]; (iv) *borated concrete wall* for neutron shielding.

The radiation fluxes produced by direct drive target explosions taken from the ARIES project [27] are used as input data for our calculations. For HiPER Demo, the spectrum of a 154 MJ target yield was selected. For HiPER Experimental and Prototype facilities, the target yields were rescaled to 20 and 50 MJ, respectively, see Ref. [10].

The most significant contributions are due to fusion neutrons (~71% of target yield), energetic burn products (H, D, T, He) and debris (D and T ions from non-burnt plasma and low energy He). In total ions carry nearly 27% of the total energy released by the explosion, while the remaining energy is carried by X-rays (1–2%). Neutron effects on the first wall can be considered negligible as compared to those of ions and X-rays, which result in a much higher power density in the irradiated material.

Accurate calculation of the thermomechanical response of the FW to irradiation requires a proper knowledge of the temporal and depth distribution of the deposited energy [28]. In

Figure 2 a) we observe how for HiPER Demo, X-rays arrive almost immediately (~22 ns) after the ignition, followed by fast burn products (D, T and He at ~0.65 µs) and debris ions (slow D,T and He at ~1.5 µs). For HiPER Prototype the ion arrival follows the same pattern as for Demo but with lower energy fluences (shots of 50 MJ instead of 154 MJ). For HiPER Experimental (smaller chamber) the irradiation arrival is quicker (~17 ns for X-rays, ~0.36 µs for fast D, T and He and ~1,1 µs for slow D,T [28]) with lower fluences (shots of 20 MJ instead of 154 MJ). Therefore, the energy deposition in the FW takes place within times lower than 10 µs. Detailed information about the mean energy, pulse duration, implantation range and energy fluence associated to each species for the different scenarios of HiPER is illustrated in Table 2.

The calculated depth distribution of the deposited energy in a W first wall for HiPER Demo is shown in

Figure 2 b). These calculations were done for: (i) ions with the aid of SRIM [29] and taking the stopping power corresponding to the average energy of each bin of ions on W; (ii) X-rays with the appropriate absorption coefficients for X-rays in W in the spectral region of interest [30]. It is important to note that more than 50% of the energy carried by X-rays and ions is deposited in the first µm of the FW and around 90 % is deposited in the first 5 µm. The rest (10 %) is deposited within depths up to 100 µm by the most energetic ions (> 1 MeV). According to these data, the thickness of the FW must be at least 100 µm in order to stop all the ions before they reach the substrate.

Table 2: Energy, pulse duration, depth range and energy fluence for the different irradiation types in the scenarios of HiPER. Radiation spectra from ARIES project [27].

| Radiation form | | Energy (%) | Mean Energy (MeV) | Pulse width (ns) | Mean Depth Range (μm) | Fluence (J/cm2) / | | |
|---|---|---|---|---|---|---|---|---|
| | | | | | | Exp. | Proto. | Demo |
| Burn | H | 0.4% | 2.55 | 500 | 10 | 0.02 | 0.03 | 0.1 |
| | D | 3.1% | 3.46 | 600 | 20 | 0.2 | 0.3 | 0.90 |
| | T | 2.8% | 2.66 | 600 | 10 | 0.2 | 0.3 | 0.8 |
| | He | 6.4% | 3.81 | 600 | 1.8 | 0.4 | 0.6 | 1.9 |
| Burn Total | | 12.7% | 3.40 | 600 | 8 | 0.8 | 1.2 | 3.7 |
| Debris | H | 0.1% | 0.09 | 1500 | 0.2 | 0.00 | 0.01 | 0.02 |
| | D | 5.8% | 0.14 | 2000 | 0.4 | 0.4 | 0.5 | 1.7 |
| | T | 7.2% | 0.19 | 2000 | 0.5 | 0.5 | 0.7 | 2.1 |
| | He | 0.9% | 0.23 | 1500 | 0.2 | 0.05 | 0.1 | 0.25 |
| Debris Total | | 14.4% | 0.17 | 2000 | 0.5 | 0.9 | 1.3 | 4.0 |
| Ions Total | | 27.1% | 0.29 | 3000 | 4 | 1.7 | 2.5 | 7.7 |
| X-rays | | 1.4% | 0.007 | 0.1 | 2 | 0.1 | 0.1 | 0.4 |
| Neutrons | | 70.8% | 12.4 | 60 | - | - | - | - |

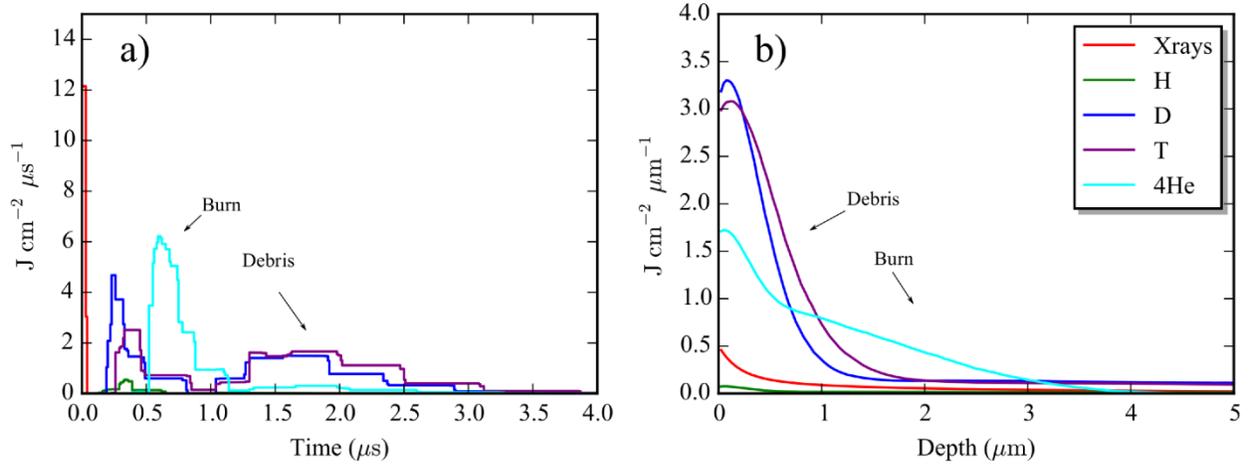

Figure 2: a) Power per unit area as a function of time and b) energy fluence as a function of depth in a W first wall for the HiPER Demo facility.

## 3. Thermomechanical Model

A schematic overview of the system used for the thermo-mechanical calculations is depicted in Figure 3. In this figure the irradiation pulse reaches the chamber, which is protected by a tungsten FW. Heat is transferred through the substrate to the substrate/coolant interface. The heat is finally removed by the LiPb coolant for HIPER Prototype and Demo or by the borated concrete shielding in the case of HIPER Experimental.

The finite element solver Code Aster [31] was used for the calculations. This software allows varying the physical parameters and mechanical properties as a function of temperature. In our particular case, the dependence of material properties on temperature was implemented using the values reported in the ITER material handbook [32, 33].

The 2D wall geometry was meshed using the Salome Platform [34]. In order to achieve a detailed estimation of the temperature gradients and local stresses, the mesh was refined near the irradiated surface (FW surface) and at the interface between the FW and the substrate. The reactor wall was supposed to be adiabatic except at the surface in contact with the coolant. The mechanical boundary conditions were fixed allowing for axial displacement of the FW surface. The thermomechanical effects appear as a consequence of the deposition of energy in the first wall. Atomistic effects such as sputtering, production of radiation-induced defects and/or changes in the chemical composition of the FW are not taken into account.

For the calculation of the mechanical response a bilinear elasto-plastic material model under the quasi-static deformation approach is used. The tungsten FW thickness is 1 mm. The substrate is assumed to be an oxide dispersion strengthened reduced activation ferritic/martensitic (ODS-RAFM) steel [26] which is able to operate at temperatures as high as 925 K. The selected thickness is 1 cm [9]. The coolant is eutectic LiPb (Pb17 at.%), which operates between 600 K and 800 K having a convection coefficient of ~10 kWm$^2$K [9].

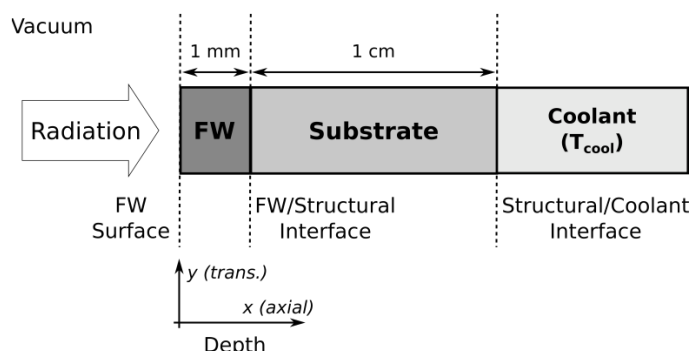

Figure 3: Sketch of the reactor wall. The W first wall is attached to the substrate. The substrate is in contact with the borated concrete in HIPER Experimental or with the coolant in HIPER Prototype and Demo.

## 4. Thermomechanical response calculations

### 4.1. Radiation-induced thermal response

Irradiation energy leads to an increase in the temperature of the FW. First the FW surface is heated during the irradiation arrival. Then the heat is transmitted through the FW to the substrate, and from the substrate to the coolant. In the Experimental case, with a limited number of shots, the FW will recover its initial temperature (300 K) shortly after irradiation. However in the Prototype and Demo cases, operating in continuous pulsed modes (1 and 10 Hz), the FW is heated during each pulse, until a steady state is reached. To characterize the steady state we define the base temperature as the temperature at the end of the pulse.

Figure 4 shows the base temperature as a function of time for the Prototype and Demo facilities. For the operation of the plant we have to consider two different regimes: the "reactor start-up" corresponding to operation times lower than ~6 s for Prototype (1 Hz, 50 MJ) and ~60 s for Demo (10 Hz, 154 MJ) and the "steady state regime" for longer times. In the case of Prototype and Demo, the initial base temperature will be of 600 K, corresponding to the temperature of the coolant. For Prototype the temperature difference between the start-up and the steady state is of only 10 K, up to 610 K in the FW surface, and will have a minor impact on the behaviour of the FW. However for the Demo case, with higher energies, the FW surface, will heat up to 934 K. Due to the high thermal conductivity of W, the temperature gradient in the FW is low and the temperature at the FW/substrate interface is 928 K. This means the substrate at the interface slightly above the operational temperature of the ODS-RAFM (925 K), which can lead to performance problems and in the joining. Finally, the temperature along the substrate decreases down to 680 K in the substrate/coolant interface for Demo.

The temperature evolution of the FW during each pulse is marked by the irradiation of X-ray, burn and debris ions, which induce three temperature peaks. In Figure 5 a), we show the temperature evolution for the Experimental facility during an experimental pulse. In the FW surface (0 µm), the temperature evolution coincides in time with the radiation arrival. Before irradiation, the FW surface is at a room temperature (300 K). With the arrival of X-rays, an initial temperature peak of 550 K is reached. Then, the temperature abruptly decreases until the arrival of ions, which takes place in two steps. First, we observe the arrival of burn ions, which heat the surface up to ~700 K. Then, the surface cools down to ~650 K and with the arrival of debris ions the temperature increases up to ~900 K. Later on, W cools down to room temperature, thus, the maximum calculated temperature for the Experimental facility is ~900 K, well below the melting point for W (3700 K [33]). Since the energy from X-rays and ions is deposited in the first microns (see

Figure 2 b), the temperature drops fast with increasing depth underneath the FW surface. For depths ≥10 µm only the arrival of ions contribute to the temperature enhancement. We also observe that for higher depths, the temperature peak appears later (due to heat conduction), having a temperature peak at ~$10^{-6}$ s for the FW surface (0 µm) and at ~$10^{-4}$ s for depths of 100 µm.

The temperature evolution in HiPER Prototype and Demo is similar to that in the Experimental facility. The main differences are the higher irradiation fluences (see Table 2) and the fact that continuous irradiation leads to a steady state regime. Figure 5 b) and c) show the temperature evolution during a pulse once the steady state regime is reached for HiPER Prototype and Demo. The temperature at the FW surface notably increases after the arrival of radiation, especially at depths ≤ 10 µm where the major part of the energy is deposited. We observe the three temperature peaks that appear during the pulse, the first one related to the X-rays, the second one due to the burn ions and the third one due to debris ion deposition. The maximum temperature is ~1400 K and ~3400 K for HiPER Prototype and Demo, respectively, which are lower than the melting point of W (3700 K [33]).

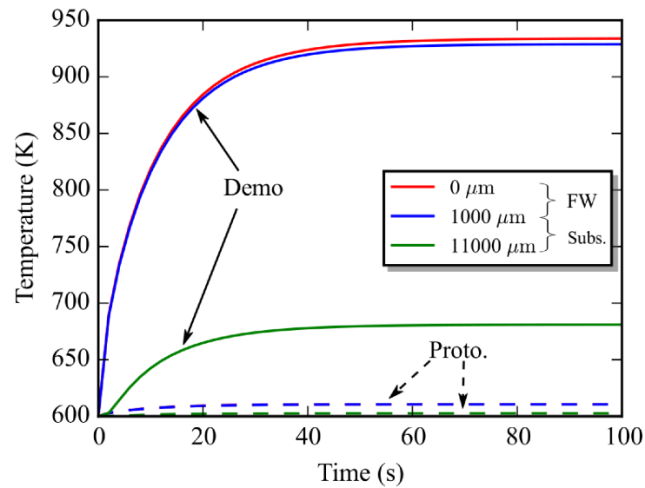

Figure 4: Calculated base temperature at the FW surface (red line), FW/substrate interface (blue line), and substrate/coolant interface (green line) as a function of time during the start-up for the HiPER Prototype (dashed line) and Demo (solid line) facilities.

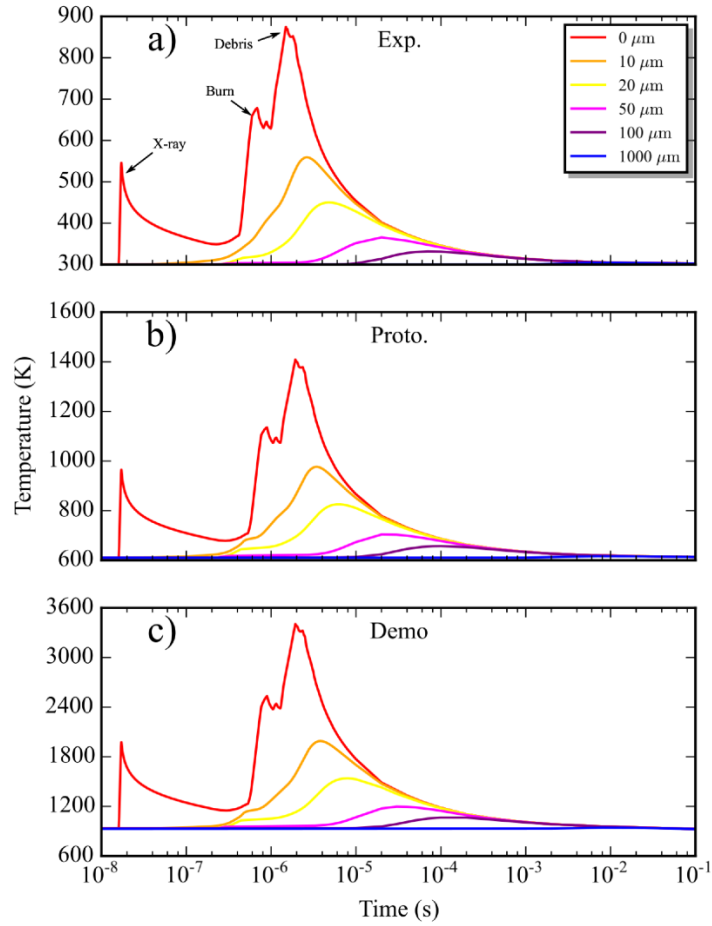

Figure 5: First wall (FW) temperature as a function of time at different depths underneath the FW surface for a) HiPER Experimental (first pulse), b) HiPER Prototype (steady state) and c) HiPER Demo (steady state).

### 4.2. Mechanical response of the W first wall

Despite the temperature never exceeds the melting point, it affects the yield stress, that decreases from around 1.2 GPa at room temperature too below 100 MPa at more than 2000 K [32]. Thus, the stresses induced by the thermal loads easily exceed the yield stress resulting in plasticization. Continuous cyclic load may lead to damage, in particular, the formation of deleterious cracks with unacceptable consequences for the FW material. We can get a first approximation to damage analysing the irradiation fluence. For the case of X-rays, the energy fluences are 0.09, 0.13 and 0.4 J/cm$^2$ for Experimental, Prototype and Demo, respectively. (see Table 2). These values are below 1 J/cm$^2$ [13] reported for damage. For the case of ion irradiation burn and debris have similar energy fluences but arrive at different times (see Table 2), with shorter pulses in the case of burn ions, which leads to more detrimental effects. We introduce the effect of pulse duration with the heat flux factor, $F_{HF}$, and for burn ions[1], we obtain $F_{HF}$ = 10.4, 15.4 and 47.3 MWm$^{-2}$s$^{0.5}$ for Experimental, Prototype and Demo. Comparing these values to the $F_{HF}$ damage threshold ~17.7 MWm$^{-2}$s$^{0.5}$ [13], we observe that HiPER Experimental is below the damage threshold. However, the values obtained for HiPER Prototype and Demo are close to or above the damage limit.

---

[1] If we calculate $F_{HF}$ considering all ion irradiation arriving in 3 µs, we have similar values of 9.7, 14.4 and 44.4 47.3 MWm$^{-2}$s$^{0.5}$ for Experimental, Prototype and Demo respectively.

In order to understand possible damage in the FW, in this section we focus on the calculation of the thermomechanical response of the first wall. The temperature enhancement during each pulse generates cyclic stresses at the FW. The origin of the stresses is due to the geometrical restrictions of the spherical chamber and the expansion of the materials associated to the temperature enhancement. Therefore, we observe an axial displacement characterized by the strains ($\epsilon_{xx}$), accompanied by a transverse constraint observed in the stresses ($\sigma_{yy}$).

In Figure 6 a), the transverse stress ($\sigma_{yy}$) and axial strain ($\epsilon_{xx}$) for the HiPER Experimental facility are depicted. During the pulse arrival (up to ~$10^{-5}$ s) the material heats up, leading to an axial expansion (positive $\epsilon_{xx}$) and transverse compression (negative $\sigma_{yy}$). In the FW surface, three peaks can be noted during the pulse. The first one is associated to the X-ray irradiation, the second one to the burn ions and the third one to debris ions. The effects of irradiation are less pronounced for increasing depth, being negligible for depths ≥100 μm. After the pulse, for times longer than $10^{-5}$ s, the material cools down, the surface retracts (decrease in $\epsilon_{xx}$) and a permanent plastic deformation is observed. In order to compensate for the axial plastic deformation, tensile transverse stresses appear (positive $\sigma_{yy}$) reaching up to ~650 MPa. Regarding the effects of irradiation below the surface, we observe only one peak for depths ≥10 μm, in contrast with the three peaks at the surface. The peak turns out less pronounced with increasing depth, being negligible for more than ~100 μm. We also observe that at higher depths the peak appears later, at around $10^{-6}$ s at the FW surface (0 μm) and $10^{-4}$ s at depths of 100 μm. At the end of the pulse, the deformation ($\epsilon_{xx}$ at the end of the pulse) is irreversible at the surface (0 μm) while at higher depths (≥ 10 μm) it decreases to negligible values.

The stresses and strains for Prototype (Figure 6 b) and Demo (Figure 6 c) scenarios differ from those of the Experimental scenario, due to three reasons: (i) operation in the steady state regime (ii) the base temperature is higher in the prototype and demo facilities and (iii) the energy per pulse is higher. Due to steady state operation, the stresses and strains before the pulse are the same as after the pulse. Temperature variation makes the FW to undergo compressive and tensile stresses. Before the pulse, the FW the material is in a tensile state, during the pulse, the surface expands axially and compresses transversely, and after the pulse recovers its initial tensile state.

In order to better understand the effect of the base temperature on the FW, it is interesting to observe the stress and strain profiles (Figure 7). We show the profiles of maximum strains and minimum stresses reached during the pulse[2] ($\epsilon_{pulse}$, $\sigma_{pulse}$) and the base values at the end of the pulse ($\epsilon_{base}$, $\sigma_{base}$) when the material recovers the base temperature. In Figure 7 a), we observe that $\sigma_{base}$ is large in the FW surface, with stresses ~GPa, quickly decreasing in the first microns defining the plastic region. The plastic region is ~5 μm for the Experimental facility, ~10 μm for the Prototype facility and ~100 μm for the Demo facility. Beyond the plastic region we observe a plateau that reaches the FW/substrate interface. This plateau appears as a consequence of the temperature mismatch between the base temperature and the room temperate. Therefore the plateau is null ($\sigma_{yy}$=0) for the Experimental (operating at room temperature), is low in the Prototype scenario ($\sigma_{yy}$~15 MPa) and much higher for Demo ($\sigma_{yy}$~300 MPa). At the FW/substrate interface a change from tensile stresses in the FW to compressive stresses in the substrate takes place for Prototype and Demo. Such a change stems from the differences in the thermal expansion coefficients of W and the substrate (larger for the ODS-RAFM steel than for W). During the pulse ($\sigma_{pulse}$), the stress becomes negative (compression stress) in the FW surface due to pulse heating. For higher depths (>100 μm) the effects of pulse heating are lower and $\sigma_{pulse}$ approaches $\sigma_{base}$, while in the substrate the stresses remain almost constant ($\sigma_{pulse}$ ~ $\sigma_{base}$).

In Figure 7 b), $\epsilon_{base}$ shows the plastic strains in the FW at the end of the pulse, when the FW cools down to the base temperature. The $\epsilon_{base}$ decreases with the depth having a plateau at the end of the plastic region. Then the $\epsilon_{base}$ has a constant value which is null in the case of the Experimental facility and lower values in Proto and Demo ($\epsilon_{xx}$ ~5·$10^{-5}$, and $\epsilon_{xx}$~2·$10^{-2}$). While in the Experimental facility the $\epsilon_{base}$ also remains null through the substrate, there is an elastic strain after the FW/substrate interface due to the material mismatch in Proto and Demo. During the pulse ($\epsilon_{pulse}$), the strain increases reaching the maximum values close to the FW surface. In Table 3 we summarize the maximum expected temperature, stress and strain at the FW surface in the different

---

[2] First pulse in the Experimental scenario and a pulse during steady state regime in the Prototype and Demo scenarios.

scenarios.

According to these data, fatigue will limit the lifetime of the FW. The maximum number of cycles ($N_f$) that a tungsten FW can withstand without failure can be calculated by means of equation $\Delta \varepsilon_p = 34 \cdot N_f^{-0.46}$ [17], where $\Delta \epsilon_p$ is the plastic strain range reached during the cycle load and $N_f$ the number of cycles for fatigue. $\Delta \epsilon_p$ can be calculated from the difference between the maximum strain reached during the pulse ($\epsilon_{pulse}$) and the permanent strain at the end ($\epsilon_{base}$). On these basis the lifetime of the Experimental facility for $\Delta \epsilon_p = 4.5 \cdot 10^{-3}$ is ~270 millions of cycles. The expected number of experiments to be carried out in this facility, assuming five shots per day during 5 years is 1800. Therefore, tungsten FW will withstand the operation conditions for which the HiPER Experimental facility is designed. For the Prototype facility, the plastic strain range is ~$10^{-2}$, allowing for ~50 000 000 cycles. Cracks would appear after 14000 hours (580 days) of operation. For Demo, the plastic strain range is ~$6 \cdot 10^{-2}$, allowing for ~1 000 000 cycles. That means that cracks will appear at the FW surface after just 28 hours of operation. This is a drastic limitation in the lifetime of the Demo, because the cracks can reach depths of up to 100 microns (see Section 4.3). Radiation damage is expected to further penetrate through the cracks, making them grow. They may eventually reach the substrate with fatal consequences (see next section).

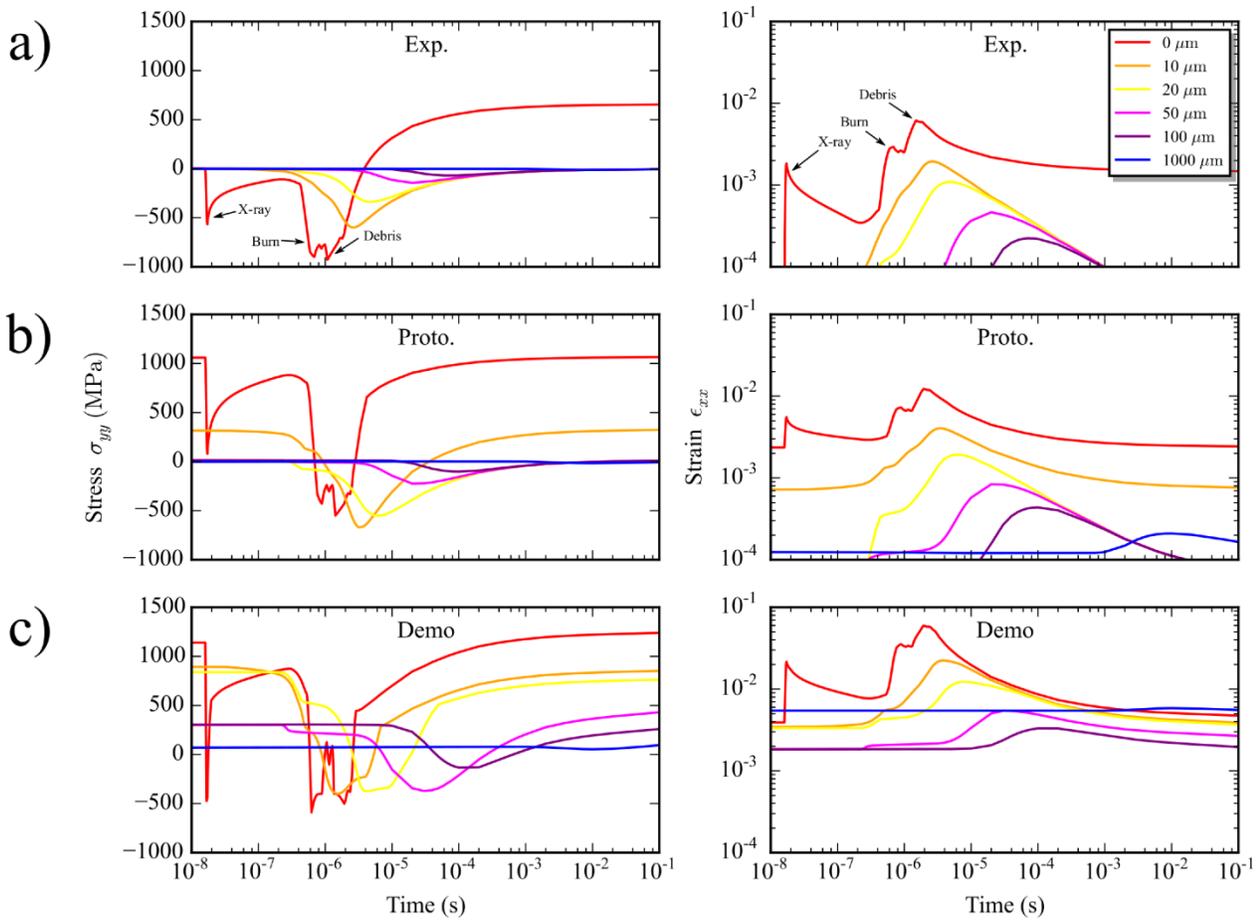

Figure 6: Transverse stress (left) and axial strains (right) as a function of time for different depths. a) HiPER Experimental (first pulse), b) HiPER Prototype (steady state) and c) HiPER Demo (steady state).

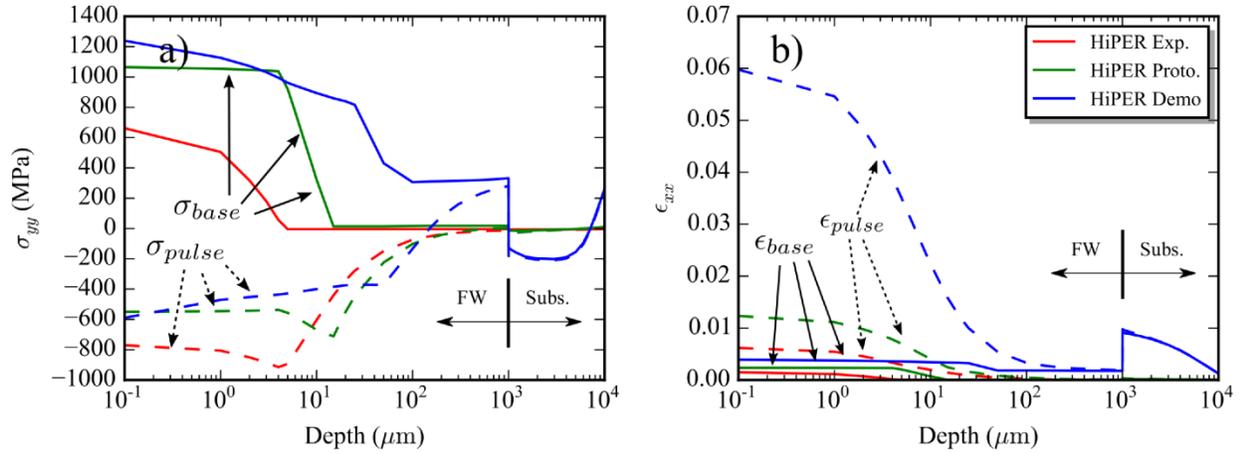

Figure 7: Transverse stress (a) and axial strain (b) profiles as a function of depth for HiPER Experimental, Prototype and Demo facilities. The stresses and strains at the end of pulse ($\sigma_{base.}$, $\epsilon_{base.}$) and during the pulse ($\sigma_{pulse.}$, $\epsilon_{pulse}$) are denoted by continuous and dashed lines, respectively.

Table 3: Main thermomechanical results for the tungsten FW surface in the different HiPER scenarios.

| HiPER Scenario | Max. Temperature (K) | | Max. Stress (MPa) | | Max. Strain | | Plastic Region (µm) |
|---|---|---|---|---|---|---|---|
| | $T_{pulse}$ | $T_{base}$ | $\sigma_{pulse}$ | $\sigma_{base}$ | $\epsilon_{pulse}$ | $\epsilon_{base}$ | |
| Experimental | 875 | 300 | -440 | 650 | 6·10⁻³ | 1.5·10⁻³ | 5 |
| Prototype | 1410 | 610 | -550 | 1050 | 1.2·10⁻² | 2.5·10⁻³ | 10 |
| Demo | 3400 | 934 | -770 | 1250 | 6·10⁻² | 4·10⁻³ | 100 |

### 4.3. Tungsten FW thickness and crack propagation

Several aspects need to be considered when estimating the required W thickness: (i) the stopping of ions in order to keep them away from the substrate (≥100 µm as discussed in Section 2), (ii) the accommodation of the stresses generated in the FW, which requires a thickness ~100 µm in order to accommodate the plastic stresses (see Figure 7), and (iii) crack propagation along the FW, to prevent undesired growth up to the substrate.

Cracks are induced by cyclic loading caused by ion irradiation. The thermal load from ion irradiation is responsible for expansion and plasticization of W under large compressive stresses. Then, after the heat deposition, W cools down and reverts the previous expansion, leading to tensile stresses in the FW surface. If there are cracks in the FW surface, they will propagate in the axial direction under the transverse tensile stress. This leads to a crack opening mode (mode I) that can be characterised with the stress intensity factor $K_I$ [35]. On these bases, the crack will propagate if $K_I$ in the crack tip is higher than the fracture toughness ($K_{IC}$) of W, which is ~5 MPa·m$^{1/2}$ for temperatures ~300 K and ~10 MPa·m$^{1/2}$ for higher temperatures [36].

This situation is illustrated in Figure 8, which shows the evolution of a crack with a length of 100 µm during the pulse duration for the Demo facility. It is observed that during the first 2 µs, W is heated, plasticizes and develops transverse compressive stresses. For times ≥ 10 µs the W surface cools down, the material in the crack edges retracts leaving an open crack with the crack tip under tensile stress.

We study how the crack propagation is affected by the crack length and the separation between cracks. The origin of the stresses in the crack tip is the plastic strain, which needs to be accommodated when the material cools down after the pulse. If the crack length is small, most of the plastic strains induce a tensile state, which leads to high stress concentration in the crack pit (high $K_I$). When the crack grows, the plastic strains lead to the opening of the crack, the material relaxes and the stress concentration in the crack tip is lower (low $K_I$). Similarly when the separation between cracks is large, all the plastic deformation between the cracks, need to be compensated by the tensile stresses, which pile up in the crack tip (high $K_I$). When the cracks are closer, the total plastic deformation is reduced, leading to lower stresses and $K_I$.

Figure 9 (a) shows $K_I$ as a function of depth for the different HiPER scenarios for a constant separation between cracks of 1000 µm. As expected, the $K_I$ factor decreases with depth. The maximum $K_I$ is 2, 5 and ~8, MPa-m$^{1/2}$ for HiPER Experimental, Prototype and Demo facilities. Assuming $K_{IC}$ to be ~ 5 MPa-m$^{1/2}$, cracks will not propagate in the Experimental facility. For the Prototype, the effect of cracks will be limited to isolated cases with depths of a few microns. For Demo, Figure 9 (b) shows that the stress intensity factor in opening mode ($K_I$) is higher than the fracture toughness ($K_{IC}$) for cracks separated more than 200 µm. Therefore, it is expected the appearance of cracks up to separations of ~200 µm are attained, which means a crack density of ~2500 cracks/cm$^2$. These cracks will propagate up to lengths of ~50-100 µm.

Based on these results a FW thickness ≥ 100 µm would correctly stop ion irradiation, accommodate temperature transients, stresses and limit crack propagation to the FW. The thickness of the tungsten FW is very relevant in order to address technological problems, such as the joining between the W and the steel [14, 37, 38], assessing it to withstand the stresses that appear in the FW/substrate interface. We propose a FW thickness of 1 mm as a conservative approach for the design of the FW. Anyway, more experimental validation of materials is necessary since tungsten damage threshold and fracture toughness can be severely affected by irradiation. Atomistic effects such as He accumulation, create additional stress concentration favouring crack appearance [39]. In this case, the high temperatures during the shot could help to release helium and reduce this effect [40] but still intensive experimentation is necessary.

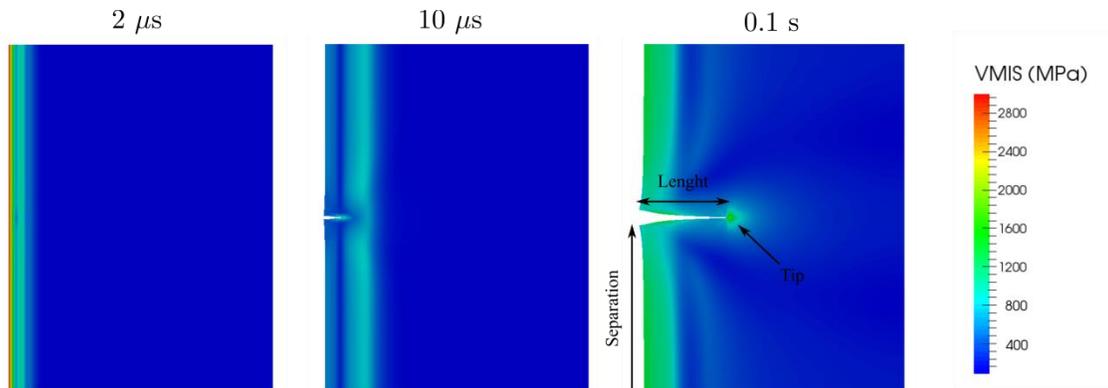

Figure 8: Evolution of a crack of length 100 µm and spacing of 400 µm in Demo reactor. The figure is deformed with a scale factor of 10x.

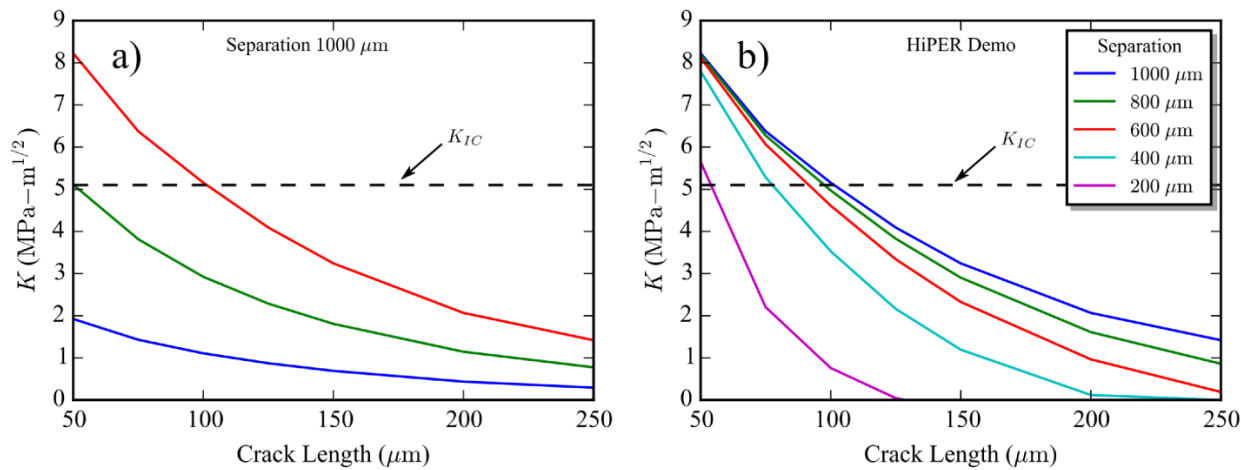

Figure 9: (a) Stress intensity factor in mode I ($K_I$) as a function of the crack length for HiPER Experimental (blue line), Prototype (green line) and Demo (red line) facilities with isolated cracks separated by 1000 μm, (b) Stress intensity factor in mode I ($K_I$) as function of the crack length for the Demo reactor with isolated cracks separated by 1000 μm (blue line), by 800 μm (green line), by 600 μm (red line), by 400 μm (light blue line), and by 200 μm (purple line) has to be below the fracture toughness $K_{IC}$ (black dashed line).

## 5. Conclusions

A systematic study on the thermomechanical response of a tungsten first wall in the different HiPER scenarios to realistic irradiation conditions is presented. The energy carried by ions and X-rays (~25% of the total shot energy) is deposited in the first 100 μm of tungsten during times of ~4 μs. During irradiation three temperature peaks appear, corresponding to pulses of X-rays, burn ions and debris ions generated with every target implosion and ignition. As a result, the FW expands until the irradiation ceases and the material cools down, leading to a tensile state due to the appearance of a plastic region. The plastic region affects the first microns of the FW.

Fatigue appears due to the cyclic nature of the irradiation. The lifetime of the FW in the different HiPER scenarios is limited by the fatigue loading. On these bases, an Experimental facility could operate with a tungsten first wall, thanks to the employed low energy targets and the limited demand of shots in such a facility. In the case of a full scale power plant (Demo), damage will appear in the FW. We encounter several problems: the peak temperature during each pulse is near the melting temperature, the energy fluence is above reported threshold damage values and cracks would appear and propagate shortly after the beginning of operation. In the case of a reactor operating with relaxed conditions (Prototype), the FW will withstand the irradiation conditions only for a short term (days). Special attention should be paid to long term operation due to fatigue problems, the expected appearance of cracks and the erosion of the FW. Regarding the thickness of the tungsten FW, the main limiting factor is crack propagation. We estimate that a tungsten FW with a thickness of ~100 μm is enough to fulfil its protection role. Such a thin FW can be fabricated by affordable methods such as PVD, offering different joining possibilities.

Since operation for full scale reactors (Demo) with a tungsten FW appears unfeasible, some scientific and technological research must still be done. This requires experiments at high temperatures under realistic irradiation conditions to better characterize the effects of irradiation and develop strengthen materials.

## 6. Acknowledgments

This work is funded by the project ENE2015-70300-C3-3-R (Radiafus-4) of Spanish MINECO and by the

Thermo-mechanical behaviour of a tungsten first wall in HiPER laser fusion scenarios             14[23] J. W. Coenen et al., "Melt-layer ejection and material changes of three different tungsten materials under high heat-flux conditions in the tokamak edge plasma of TEXTOR," *Nucl. Fusion*, vol. 51, no. 11, p. 113020, 2011.
[24] T. Loewenhoff et al., "Evolution of tungsten degradation under combined high cycle edge-localized mode and steady-state heat loads," *Phys. Scr.*, vol. T145, p. 014057, Dec. 2011.
[25] A. R. Raffray, "Threats, design limits and design windows for laser IFE dry wall chambers," *J. Nucl. Mater.*, vol. 347, no. 3, pp. 178–191, 2005.
[26] N. Baluc et al., "Status of R&D activities on materials for fusion power reactors," *Nucl. Fusion*, vol. 47, no. 10, pp. S696–S717, Oct. 2007.
[27] "ARIES Web Site -- ARIES-IFE Documents." http://aries.ucsd.edu/ARIES/WDOCS/ARIES-IFE/.
[28] J. Alvarez, D. Garoz, R. Gonzalez-Arrabal, A. Rivera, and M. Perlado, "The role of spatial and temporal radiation deposition in inertial fusion chambers: the case of HiPER," *Nucl. Fusion*, vol. 51, no. 5, p. 053019, May 2011.
[29] J. F. Ziegler, M. D. Ziegler, and J. P. Biersack, "SRIM – The stopping and range of ions in matter (2010)," *Nucl. Instrum. Methods Phys. Res. Sect. B Beam Interact. Mater. At.*, vol. 268, no. 11–12, pp. 1818–1823, Jun. 2010.
[30] J. H. Hubbel and S. M. Seltzer, "NIST: X-Ray Mass Attenuation Coefficients." http://www.nist.gov/pml/data/xraycoef/index.cfm.
[31] EDF R&D, "Manuel d'utilisation: introduction au Code_Aster. Electricité de France." http://www.code-aster.org/V2/doc/v12/fr/man_u/u1/u1.02.00.pdf.
[32] J. W. Davis and P. D. Smith, "ITER material properties handbook," *J. Nucl. Mater.*, vol. 233–237, Part 2, pp. 1593–1596, Oct. 1996.
[33] "ITER Material Properties Handbook, ITER Document No. S 74 RE 1." .
[34] "SALOME Platform." http://www.salome-platform.org/.
[35] T. L. Anderson, *Fracture Mechanics: Fundamentals and Applications,* Third Edition. Boca Raton, FL: CRC Press, 2005.
[36] B. Gludovatz, S. Wurster, A. Hoffmann, and R. Pippan, "Fracture toughness of polycrystalline tungsten alloys," *Int. J. Refract. Met. Hard Mater.*, vol. 28, no. 6, pp. 674–678, Nov. 2010.
[37] J. Schlosser et al., "Design, fabrication and testing of an improved high heat flux element, experience feedback on steady state plasma facing components in Tore Supra," *Fusion Eng. Des.*, vol. 39–40, pp. 235–240, Sep. 1998.
[38] V. Barabash et al., "Armor and heat sink materials joining technologies development for ITER plasma facing components," *J. Nucl. Mater.*, vol. 283–287, Part 2, pp. 1248–1252, Dec. 2000.
[39] T. J. Renk et al., "Survivability of First-Wall Materials in Fusion Devices: An Experimental Study of Material Exposure to Pulsed Energetic Ions," *Fusion Sci. Technol.*, vol. 61, no. 1, pp. 57–80, 2012.
[40] S. B. Gilliam et al., "Retention and surface blistering of helium irradiated tungsten as a first wall material," *J. Nucl. Mater.*, vol. 347, no. 3, pp. 289–297, Dec. 2005.